\documentclass{article}

\PassOptionsToPackage{numbers}{natbib}

\usepackage[preprint]{neurips_2024}

\usepackage[utf8]{inputenc} 
\usepackage[T1]{fontenc}    
\usepackage{hyperref}       
\usepackage{url}            
\usepackage{booktabs}       
\usepackage{amsfonts}       
\usepackage{nicefrac}       
\usepackage{microtype}      
\usepackage{xcolor}         
\usepackage{amsmath}
\usepackage{amsthm}
\usepackage{nicematrix}
\usepackage{algorithm}
\usepackage{algpseudocode}
\usepackage{enumitem}
\usepackage{graphicx}
\usepackage{wrapfig}
\usepackage{lipsum}
\usepackage{soul}

\theoremstyle{plain}
\newtheorem*{theorem*}{Theorem}

\title{Contribute to balance, wire in accordance: Emergence of backpropagation from a simple, bio-plausible neuroplasticity rule}

\author{%
Xinhao Fan$^{1*}$ \quad Shreesh P Mysore$^{1,2*}$ \\
$^1$Department of Neuroscience \\
$^2$Department of Psychological and Brain Sciences\\
$^*$Corresponding author\\
Johns Hopkins University\\
Baltimore, MD, 21218\\
\texttt{\{xfan20, mysore\}@jhu.edu}
}

\begin{document}

\maketitle

\begin{abstract}
  Over the past several decades, backpropagation (BP) has played a critical role in the advancement of machine learning and remains a core method in numerous computational applications. It is also utilized extensively in comparative studies of biological and artificial neural network representations. Despite its widespread use, the implementation of BP in the brain remains elusive, and its biological plausibility is often questioned due to inherent issues such as the need for symmetry of weights between forward and backward connections, and the requirement of distinct forward and backward phases of computation. Here, we introduce a novel neuroplasticity rule that offers a potential mechanism for implementing BP in the brain. Similar in general form to the classical Hebbian rule, this rule is based on the core principles of maintaining the balance of excitatory and inhibitory inputs as well as on retrograde signaling, and operates over three progressively slower timescales: neural firing, retrograde signaling, and neural plasticity. We hypothesize that each neuron possesses an internal state, termed credit, in addition to its firing rate. After achieving equilibrium in firing rates, neurons receive credits based on their contribution to the E-I balance of postsynaptic neurons through retrograde signaling. As the network's credit distribution stabilizes, connections from those presynaptic neurons are strengthened that significantly contribute to the balance of postsynaptic neurons. We demonstrate mathematically that our learning rule precisely replicates BP in layered neural networks without any approximations. Simulations on artificial neural networks reveal that this rule induces varying community structures in networks, depending on the learning rate. This simple theoretical framework presents a biologically plausible implementation of BP, with testable assumptions and predictions that may be evaluated through biological experiments.
\end{abstract}

\section{Introduction}
 Backpropagation (BP) is the most widely used supervised learning algorithm for training artificial neural networks. At its core, it updates  the weights of a structured neural network in terms of the gradient of an  error function with respect to these weights, doing so via the application of the Leibniz chain rule.  BP has achieved remarkable success in diverse fields including image \cite{krizhevsky2012imagenet} and speech recognition \cite{amodei2016deep}, natural language processing \cite{vaswani2017attention}, financial prediction \cite{sezer2020financial}, and medical diagnosis \cite{rajpurkar2017chexnet}. Despite its effectiveness within machine learning, a major criticism of BP has been its lack of biological plausibility. Two key issues have been raised in the literature in this context: (1) the `weight symmetry' problem, and (2) the `update locking' problem. The weight symmetry problem refers to the requirement that the feedback connectivity matrix needs to be the exact transpose of the feedforward connectivity matrix in order to propagate error gradients accurately. The time-locking problem refers to BP's two distinct phases of feedforward and feedback information processing: respectively, the forward-flowing updates of unit activation states (or firing rates) across the layers of the network, followed by the backward-flowing updates to weights between successive pairs of layers. These two serially implemented stages of information flow necessarily require that certain parts of the network are transiently `locked', i.e., can be updated only after other parts, resulting in `locking'. As a result, learning cannot occur `online', or simultaneuously,  as the network is actively engaged in task-solving through feedforward processing. Researchers have explored various strategies to address both problems with varying degrees of success. 

With respect to the weight symmetry problem, \textit{one strategy} has been to relax the requirement for an exact transpose of the feedforward connectivity matrix. 
The feedback alignment (FA) algorithm \cite{lillicrap2016random} does this by starting with random feedback connectivity, that is shown to gradually align with the transpose of the feedforward matrix. 
The sign symmetry (SS) framework \cite{liao2016important} goes further, requiring that only the signs of the feedback connections match that of the transposed feedforward matrix.
The use of random connectivity was later shown to impair performance on larger-scale problems \cite{russakovsky2015imagenet, bartunov2018assessing}. 
In response,the weight mirror (WM) mechanism \cite{akrout2019deep} employed specific neural circuits to approximate the feedforward connectivity more accurately. 
However, it remains uncertain whether these algorithms can perform as well as BP on diverse tasks, and the biological mechanisms for such complex feedback tuning are largely unknown. \textit{Another strategy} proposes entirely different feedback computation methods from that in the original BP framework. Direct/Indirect Feedback Alignment (DFA/IFA) \cite{nokland2016direct} replaces layer-to-layer feedback connectivity with direct feedback connections from the output layer to the hidden layers, eliminating the need for layer-by-layer propagation. Global Error Vector Broadcasting (GEVB) \cite{clark2021credit} simplifies the feedback signal to a single vectorized error signal permeated across the entire network. Target Propagation (TP), a distinct family of solutions \cite{le1986learning, bengio2014auto, meulemans2020theoretical, lee2015difference, bartunov2018assessing, frenkel2021learning}, propagates the correct activation to each hidden neuron rather than the error, thereby removing the necessity for symmetry. Additionally, the concept of a separate teacher network has been proposed, wherein the teacher network tunes the feedback to be symmetrical to the feedforward connectivity. This idea has been framed within a reinforcement learning context in node perturbation (NP) \cite{lansdell2019learning}, and as a controller in deep feedback control (DFC) \cite{meulemans2021credit} and least control principle (LCP) \cite{meulemans2022least}. However, the newly proposed feedback computations in most models lack biological validation, and drastically redesigning BP feedback could lose many of its beneficial properties. A \textit{third strategy} has been to directly invoke a biological mechanism that permits synapses to transmit information backward along the axon to the soma, namely retrograde signaling. As the same synapse is used in both directions, symmetry is inherently maintained. Although some research has explored this avenue \cite{harris2008stability}, the fundamental limitation perceived with retrograde signaling in the context of BP is that it occurs on a much slower timescale compared to neural firing \cite{lillicrap2020backpropagation, oztas2003neuronal}. 
Nonetheless, given the established biological feasibility of retrograde signaling, novel strategies that could achieve congruence between its slow timescale and the rapid timescale of feedforward activity are a potentially promising (but overlooked) route to resolving the weight symmetry problem. 

With respect to addressing the update locking problem, significant efforts have been made to unify the feedforward and feedback phases into a single phase. Predictive Coding (PC) \cite{friston2005theory, whittington2017approximation} attempts to tackle this issue by introducing a second type of neuron that encodes prediction errors and influences network updates. The error neurons are only active when errors are present, allowing PC to automatically differentiate between phases. 
Similarly, Equilibrium Propagation (EP) \cite{scellier2017equilibrium, scellier2019equivalence, laborieux2022holomorphic} recognizes phases by describing the system through a unified rule of minimizing Hopfield energy, with and without the output layer clamped by an external signal. LCP \cite{meulemans2022least} integrates error assignment and system firing dynamics into a single phase of optimal control within a control theory framework. Direct Random Target Propagation (DRTP) \cite{frenkel2021learning} replaces backward error propagation with forward target activity propagation, thus maintaining only two types of forward propagation within the model. Additionally, other studies attempt to leverage more biologically plausible mechanisms such as spike-timing-dependent plasticity (STDP) to resolve this issue \cite{payeur2021burst}. 
Broadly speaking, however, there does not yet exist a widely accepted approach of rule unification for the resolution of the update locking problem. Novel strategies that, instead, use distinct rules operating simultaneously on different timescales are a potentially promising (but unexplored) alternative. 

In addition to the widely investigated weight symmetry and update locking problems, there is a third potential concern regarding the biological implausibility of BP, namely (3) the `derivative computation' problem: how can neurons compute  derivatives? Although not the main focus of research in this field, this question has been addressed implicitly in many frameworks, with partial success. In energy-based models such as PC and EP, derivatives are introduced through the first principle of energy minimization, where the network evolves according to the gradient of energy \cite{friston2005theory, whittington2017approximation, scellier2017equilibrium, scellier2019equivalence}. For more biologically realistic neurons, derivatives are embedded in the differential dynamics governing neural activity \cite{payeur2021burst}. Alternatively, derivatives can be computed by a teacher network acting as a reinforcement learning agent \cite{lansdell2019learning} or a PID controller \cite{meulemans2021credit}. Some frameworks avoid the direct computation of derivatives by replacing it with an algebraic computation. For instance, in GEVB, specific forms of activation nonlinearity are assumed \cite{clark2021credit}, while in contrastive Hebbian learning (CHL) \cite{movellan1991contrastive, xie2003equivalence} and generalized recirculation (GeneRec) \cite{hinton1987learning, o1996biologically}, the use, solely, of multiplication is a limiting case of BP, resulting in a Hebbian-like rule. The TP family goes a step further, replacing the derivative of the loss function with a target signal \cite{le1986learning, bengio2014auto, meulemans2020theoretical, lee2015difference, bartunov2018assessing, frenkel2021learning}, thereby eschewing altogether the need for the derivative computation in its setup.

Despite the numerous alternative frameworks that have been proposed, BP remains the most powerful learning algorithm for a diverse range of tasks and datasets, and is still the most widely used method for machine learning applications and developments. Some alterative frameworks can mathematically reduce to BP, but they require additional assumptions, such as a weak teaching signal in EP \cite{scellier2019equivalence} or weak feedback connections in CHL \cite{xie2003equivalence}, which lack real-world relevance and are merely mathematical constraints.
Furthermore, many frameworks attempting to solve the bio-plausibility problem of BP introduce additional requirements that may themselves be biologically implausible. Examples include the specific structure of neural circuits in DFA/IFA, WM, and PC \cite{nokland2016direct, akrout2019deep, whittington2017approximation}, as well as the widespread existence of teacher networks in NP and DFC \cite{lansdell2019learning, meulemans2021credit}. Consequently,an enduring open question remains, "what is a learning rule implementable by the brain that possesses the features of, and exhibits the achievements realized by, BP?"

Here, we propose a new framework for learning in neural networks based on the foundational principle that neurons pursue excitatory-inhibitory (E-I) balance in their inputs. This framework operates on three progressively slower timescales: neural firing for computation, retrograde signaling for credit assignment, and neural plasticity for weight updates. On the one hand, this formulation allows the system to automatically detect when to update weights, thereby resolving the (2) update locking problem. On the other, it addresses the (1) weight symmetry problem by embedding the slower time-scale retrograde signaling naturally within its three-timescale operation. Separately, by hypothesizing the existence of an internal mechanism in each neuron that measures deviations from its preferred state of E-I balance, we derive a three-factor Hebbian-like neuroplasticity rule that does not rely on derivatives, thereby resolving the (3) derivative computation problem. We prove that in layered neural networks, this framework can be mathematically reduced to BP, \textit{without any approximations}. Additionally, our framework makes several experimentally testable predictions.


\section{The Model}
\subsection{E-I balance as the central principle} 
The core conceptual foundation of our neuroplasticity framework is the maintenance of E-I balance of neurons. E-I balance refers to the equilibrium between excitatory and inhibitory inputs to neurons. 
Maintaining this balance has been shown to help direct neuronal firing toward the linear response zone, which is essential for efficient information processing and representation \cite{van1996chaos, deneve2016efficient, zhou2018synaptic}. 
Although multiple biological mechanisms have been shown to contribute to maintenance of E-I balance, including homeostatic regulation of synaptic strengths \cite{turrigiano2004homeostatic, turrigiano2008self}, alterations in the intrinsic excitability of neurons \cite{desai1999plasticity}, 
and heterosynaptic plasticity \cite{field2020heterosynaptic}, learning rules have, as yet, not incorporated this requirement directly into their formulations. Our model fills this gap.

\subsection{Neural firing}
\begin{figure}[h!]
  \centering
  \includegraphics[width=0.8\textwidth]{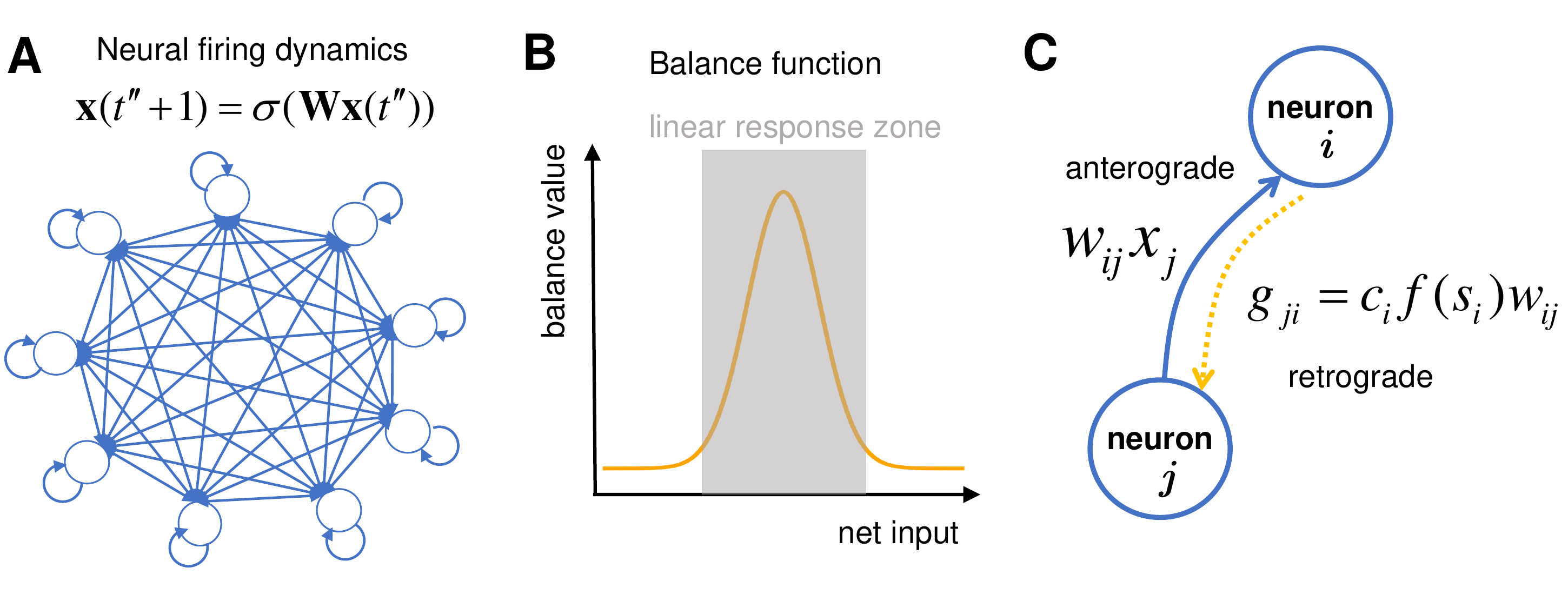}
  \caption{The model. (A) All-to-all connectivity in an ANN. (B) Bell-shaped function quantifying E-I balance of a neuron. (C) Credit distribution from postsynaptic neuron $i$ to presynaptic neuron $j$ (hypothesized to occur via retrograde signaling mechanisms, orange arrow).}
  \label{fig:model}
  \vspace{-2em}
\end{figure}

We utilize an all-to-all connected neural network to establish our model (Figure \ref{fig:model}A), which provides a high degree of generality. The firing activities of neurons are represented as an $n$-dimensional vector $\mathbf{x}(t'')$, where the $i^{\text{th}}$ element $x_i(t'')$ reflects the firing activity of the $i^{\text{th}}$ neuron at time step $t''$. The firing activity depends on the nonlinear activation function $\sigma$ and the inputs to each neuron $\mathbf{s}(t'') = \mathbf{Wx}(t'')$, where $\mathbf{W}$ is the connectivity matrix. The element $w_{ij}$ in $\mathbf{W}$ indicates the strength of the synapse received by neuron $i$ from neuron $j$. Consequently, the network dynamics and steady state (fixed points) can be described as follows:
\begin{equation}
    \mathbf{x}(t''+1) = \sigma (\mathbf{W} \mathbf{x}(t'')), \quad \mathbf{x} = \sigma (\mathbf{W} \mathbf{x})
    \label{eq:x=sigmaWx}
\end{equation}
The time $t$ here is used specifically for the dynamics of the firing activity, and is in the range of milliseconds.

\subsection{Neural credits}
In addition to the firing rate, we hypothesize that each neuron encodes a second quantity that we refer to as \textit{credit}. Credits of neurons are represented as an $n$-dimensional vector $\mathbf{c}(t')$, where the $i^{\text{th}}$ element $c_i(t')$ reflects the amount of credit possessed by the $i^{\text{th}}$ neuron at time step $t'$.

The magnitude of this quantity for each neuron represents the extent to which it has contributed to the E-I balance of its postsynaptic neurons. In turn, the degree of E-I balance for a postsynaptic neuron is measured as the deviation of the neuron's net input from the range that produces a linear response, and is quantified by a \textit{balance function} $f$: a bell-shaped function which peaks at the balanced set-point of net input, and decreases toward zero when the net input deviates from the set-point (Figure \ref{fig:model}B). With this framework in place, the credit $g_{ji}$ distributed from postsynaptic neuron $i$ to presynaptic neuron $j$ is governed by the following rule (Figure \ref{fig:model}C):
\begin{equation}
    g_{ji} = c_if(s_i)w_{ij} 
    \label{eq:g_ji}
\end{equation}
where $c_i$ is the amount of credit possessed by the postsynaptic neuron $i$; $s_i = \sum_k w_{ik} x_k$ is neuron $i$'s input and $f(s_i)$ represents how balanced it is. 

This credit redistribution rule encapsulates three underlying relationships: \textbf{(i) }postsynaptic neurons with more credits give more credits to their presynaptic neurons; \textbf{(ii)} postsynaptic neurons that are more balanced give more credits to their presynaptic neurons; and \textbf{(iii)} presynaptic neurons receive more credits if the synaptic strength is stronger. From a biological perspective, we propose that neuronal credits are encoded via biomolecules related to retrograde messengers, and distributed via retrograde signaling from postsynaptic to presynaptic neurons (see also section 5).

The dynamics of credit redistribution for each presynaptic neuron $j$ can be described as $c_j(t'+1) = \sum_i g_{ji}(t')$, with the summation occuring over all of its postsynaptic neurons $i$. Consequently, the network dynamics of credit redistribution and the corresponding fixed points can be described as:
\begin{equation}
\mathbf{c}(t'+1) = \mathbf{J} \mathbf{c}(t') = \begin{bmatrix}
f(s_1)w_{11} & f(s_2)w_{21} & \cdots & f(s_n)w_{n1} \\
f(s_1)w_{12} & f(s_2)w_{22} & \cdots & f(s_n)w_{n2} \\
\vdots & \vdots & \ddots & \vdots \\
f(s_1)w_{1n} & f(s_2)w_{2n} & \cdots & f(s_n)w_{nn}
\end{bmatrix}
\begin{bmatrix}
c_1(t') \\
c_2(t') \\
\vdots \\
c_n(t')
\end{bmatrix}
, \quad \mathbf{c} = \mathbf{J} \mathbf{c}
\label{eq:c=Jc}
\end{equation}
The matrix $\mathbf{J}$ summarizes these dynamics. Here, $t'$ is the time index specifically for the dynamics of credit redistribution, which is independent from the time index in neural firing dynamics $t''$. The time for credit redistribution is expected to be in the range of a second or a few seconds, based on the known biological features of retrograde signaling.

\subsection{Neuroplasticity rule}
With the neural firing activity and credit values defined, we propose a new three-factor plasticity rule as an extension to the two-factor Hebbian rule. In it, both the firing activity and the contribution to network E-I balance are taken into account for weight updating, which we mathematically formalize:
\begin{equation}
    \Delta w_{ij} =   c_i f(s_i) x_j, \quad \text{i.e.} \quad \mathbf{W}(t+1) = (\mathbf{c} \odot f(\mathbf{s})) \mathbf{x}^T  + \mathbf{W}(t)
    \label{eq:dW rule}
\end{equation}
with the $\odot$ symbol denoting point-wise multiplication.  

The three factors in this learning rule encapsulate the following relationships: \textbf{(a)} If the postsynaptic neuron has more credits, the synapse will be strengthened more; \textbf{(b) }If the postsynaptic neuron is more balanced, the synapse will be strengthened more. \textbf{(c)} If the presynaptic neuron has a higher firing rate, the synapse will be strengthened more. Separately, the timing for weight updates in equation \ref{eq:dW rule} is expected to operate in the range of seconds to minutes, on stable values of both firing rates $x_j$ as well as of credit values $c_i$.

\paragraph{Neurons that contribute to balance, wire in accordance.} The product term $c_i f(s_i)$, also shared in equation \ref{eq:g_ji}, represents the effective credit a postsynaptic neuron can distribute to other neurons. This is because a higher credit value possessed by a postsynaptic neuron reflects a greater potential to distribute credits to its presynaptic neurons, and a higher degree of E-I balance suggests a higher propensity to allocate those credits. Consequently, establishing connections with neurons that have higher effective credits increases the likelihood of receiving more credits in the future. Separately, a higher firing rate of a presynaptic neuron leads to the strengthening of all its synapses, which is consistent with Hebbian plasticity. Taken together, presynaptic neurons that fire will strengthen their connections to postsynaptic neurons that contribute significantly to network balance and are well-balanced themselves. We encapsulate this rule with the phrase "Neurons that contribute to balance, wire in accordance," mirroring Hebb's principle: "Neurons that fire together, wire together."

\subsection{Timescales of the model}
\begin{figure}[h!]
  \centering
  \includegraphics[width=0.85\textwidth]{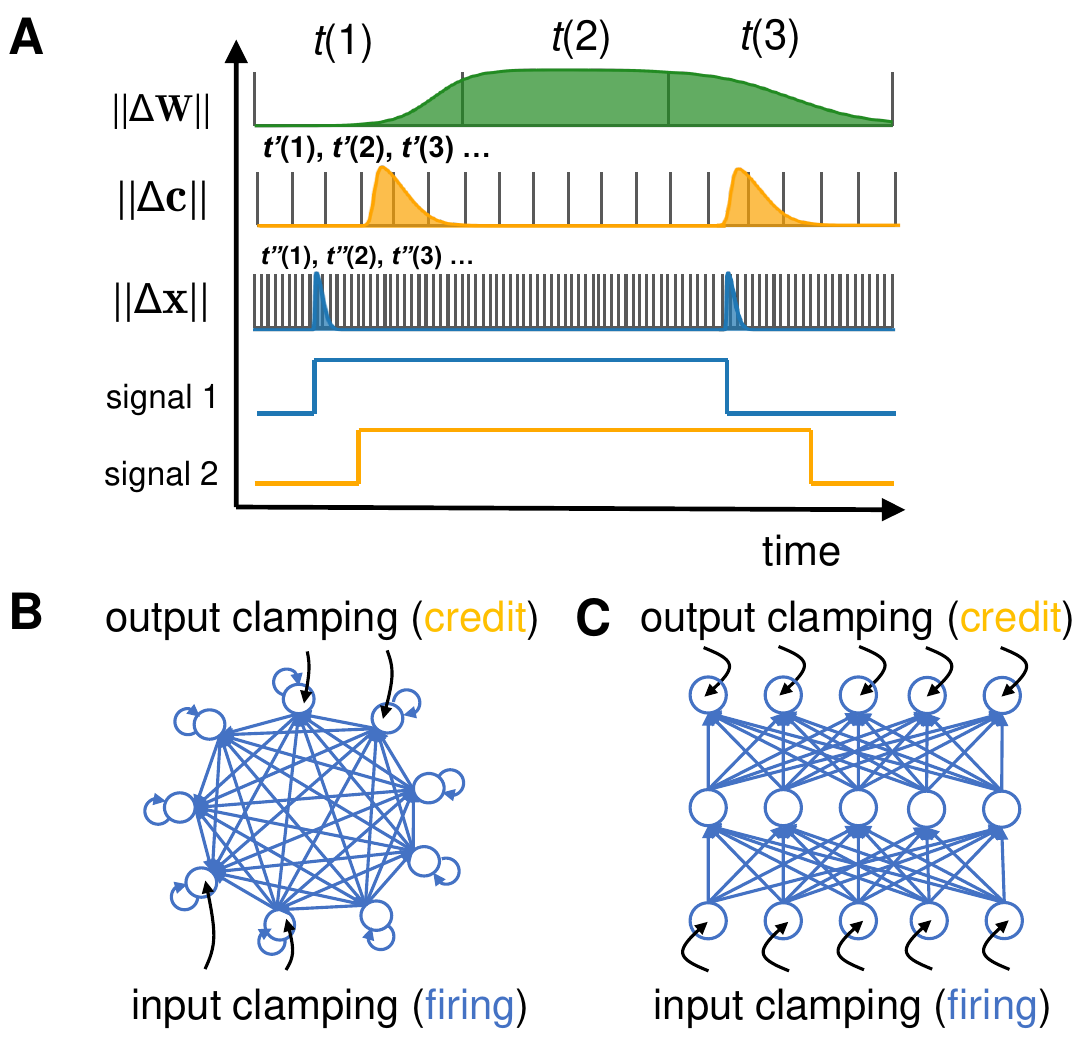}
  \caption{Three timescales of network operation. (A) Each row represents updates to the corresponding quantity; timescales are progressively faster from top to bottom. For the top three rows, each vertical line represents one timebin, and the curve represents the amplitude of the rate of updates. The area under the curve thus represents the total amount of change made to different quantities. As the quantity converges to steady-state, the corresponding curve value decreases to zero. Learning occurs with input clamping on neural firing (signal 1, blue) and output clamping on credit values (signal 2, yellow), as long as there is a sufficiently long period of overlap. (B,C) Input and output clamping in an all-to-all connected ANN, and a layered ANN, respectively.}
  \label{fig:timescale}
  \vspace{-2em}
\end{figure}

Our model operates on three sequentially slower timescales: firing activity on the millisecond scale, credit redistribution on the seconds scale, and neural plasticity from seconds to minutes. This is depicted graphically in the raster plot of figure \ref{fig:timescale}A, where each update mechanism is shown at its respective timescale using distinct time bin sizes for reference. When the system undergoes significant changes (for instance, owing to the arrival of a new input), the amplitude of curve is high. As the system stabilizes and converges, the amplitude gradually decreases to zero. Due to the hierarchical nature of these different timescales, we posit that the time it takes for one level to converge is much shorter than the update time interval of the subsequent level. Consequently, most changes due to credit update (yellow shadow under the curve) occur when network firing has already converged, and most weight changes (green shadow) occur when credit redistribution has converged. In other words, equation \ref{eq:c=Jc} is based on the equilibrium state of equation \ref{eq:x=sigmaWx}, while equation \ref{eq:dW rule} is based on the equilibrium state of equation \ref{eq:c=Jc}.

For learning to occur, the neuroplasticity rule (equation \ref{eq:dW rule}) requires both firing activity $\mathbf{x}$ and credit distribution $\mathbf{c}$ to be non-zero. However, these values can naturally converge to a zero state in the absence of external sensory signals, as $\mathbf{0} = \mathbf{J}\mathbf{0}$ and $\mathbf{0} = \sigma(\mathbf{W}\mathbf{0})$ for many activation functions such as \textit{tanh} or \textit{shifted sigmoid}. To address this, we propose that two sets of "clampings" are necessary for learning to take place. One subgroup of neurons needs to have their firing activity clamped to ensure the network converges to a non-zero state, while a different subgroup of neurons needs to have their credit value clamped. We refer to the former as input neurons and the corresponding clamping as \textit{input clamping}, and the latter as output neurons and \textit{output clamping}. We note that the term "clamping", here, is used to mean sensory inputs or error/reward signals from other neural circuits occurring for some duration of time (see also Discussion).

The time windows for input clamping (signal 1) and output clamping (signal 2) are illustrated in figure \ref{fig:timescale}A. An overlap between these two time windows is necessary because, outside this overlap, either $\mathbf{x}$ or $\mathbf{c}$ will quickly converge to zero. Additionally, the overlap must be long enough for neuroplasticity to occur. Therefore, the durations of both input and output clamping should align with the neuroplasticity timescale, which is from seconds to longer. As a result of these multiple timescales, the system automatically distinguishes between the prediction phase (only input clamping) and the learning phase (overlapped input and output clamping), thereby avoiding the need to explicitly introduce distinct phases. From a biological perspective, we hypothesize that these multiple hierarchical timescales are orchestrated by neural oscillations (see also section 5).

\subsection{Summary of the model}
In summary, our new framework operates following three update rules: \textbf{(1)} Neural firing dynamics at the fastest timescale: $\mathbf{x}(t''+1) = \sigma (\mathbf{W} \mathbf{x}(t''))$; \textbf{(2)} Credit redistribution dynamics at the intermediate timescale: $\mathbf{c}(t'+1) = \mathbf{J} \mathbf{c}(t')$; 
\textbf{(3)} Neural plasticity dynamics at the slowest timescale: $\mathbf{W}(t+1) = (\mathbf{c} \odot f(\mathbf{s}))\mathbf{x}^T + \mathbf{W}(t)$


\section{Model reduces \textit{exactly} to backpropagation in layered neural networks}
Here, we demonstrate the link between our proposed neuroplasticity framework and BP. We will first provide an intuitive understanding of how the chain rule emerges from our model setup, using derivations in an all-to-all connected neural network (Figure \ref{fig:timescale}B). We will then present a theorem proving the equivalence of both methods in layered neural networks.

Assuming the credit values for the network start with the initial value $\mathbf{c}(0)$, and following the credit redistribution rule in equation \ref{eq:g_ji}, the elements for the credits at the next step $\mathbf{c}(1)$ are therefore:
\begin{equation}
    \mathbf{c}(1) =\mathbf{J}\mathbf{c}(0), \quad \text{i.e.} \quad c_i(1) = \sum_j f(s_j) w_{ji} c_j(0)
\end{equation}
Following this line of thinking, the subsequent credit distribution can also be derived:
\begin{align}
    c_i(2) &= \sum_j f(s_j) w_{ji} c_j(1) = \sum_{j,k} f(s_j) w_{ji} f(s_k) w_{kj} c_k(0) \\
    c_i(3) &=  \sum_{j,k,l} f(s_j) w_{ji} f(s_k) w_{kj} f(s_l) w_{lk} c_l(0)  \quad ... ...
\end{align}
where an interleaved pattern of $w$ and $f$ emerges. For simplicity, we approximate $\mathbf{c}(2)$ as the final credit distribution at equilibrium. Consequently, the neuroplasticity update based on this distribution, following equation \ref{eq:dW rule}, is:
\begin{align} 
    \Delta w_{ij}  = \sum_{k,l} f(s_k) w_{ki} f(s_l) w_{lk} c_l(0)f(s_i)x_j = \sum_{k,l} c_l(0) f(s_l) w_{lk} f(s_k) w_{ki} f(s_i)x_j
\end{align}

\begin{wrapfigure}{r}{0.3\textwidth}
  \centering
  \includegraphics[width=0.25\textwidth]{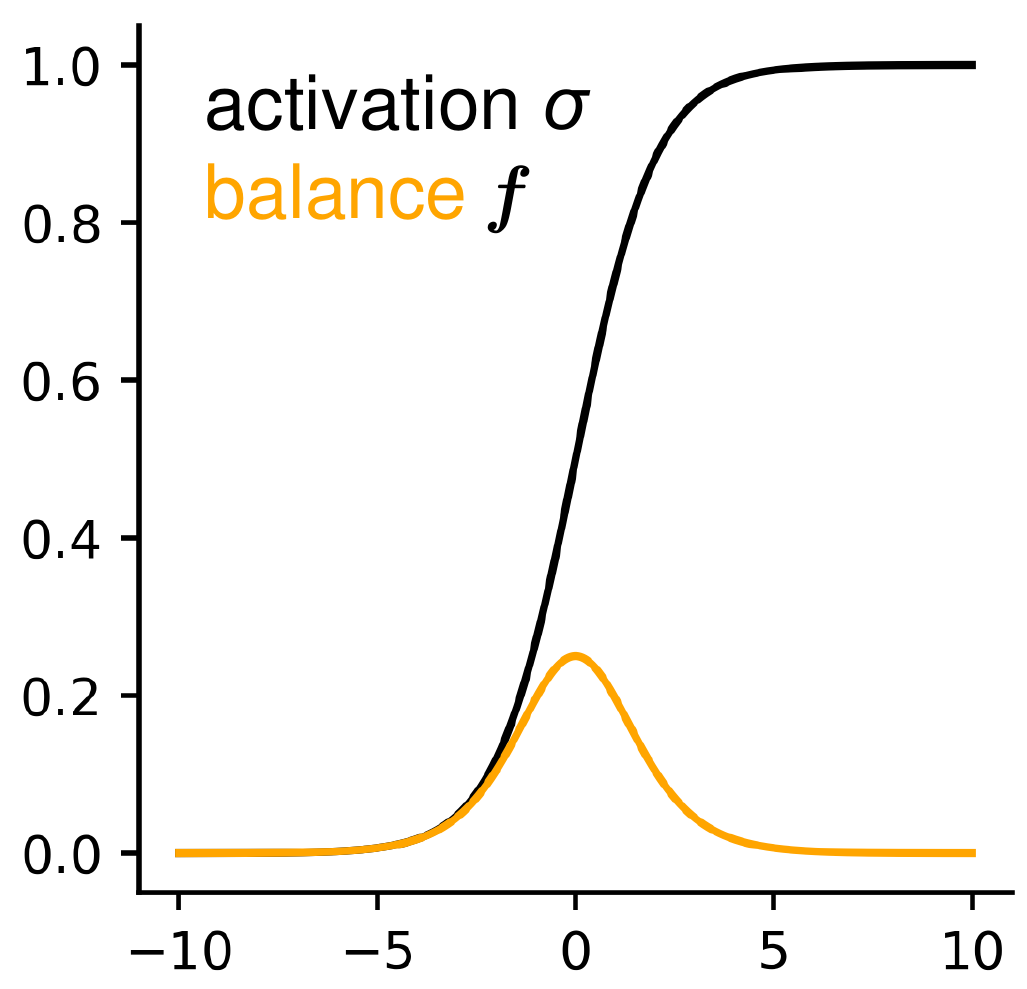}
  \caption{Choosing the balance function (orange curve) as the derivative of the activation function (black curve). }
  \label{fig:sigmoid and deri}
  \vspace{-4.1em}
\end{wrapfigure}

To further illustrate the connection between our model and gradient descent, we need to specify the form of the balance function $f$. This brings us to the \textit{key idea of linking to BP}, which relies on the chain rule and gradient computations, with our framework, which involves simple algebraic computations. Specifically, we propose that the derivative of the activation function serves as the balance function, i.e., $\partial \sigma / \partial s = f$ (Figure \ref{fig:sigmoid and deri}). 

The derivative of the activation function is a natural choice for the balance function because, for monotonically increasing activation functions that saturate at the extremes, its derivative typically forms a bell-shaped curve centered around the linear response zone. For instance, the derivative of the sigmoid function exhibits this property (Figure \ref{fig:sigmoid and deri}). For other activation functions like ReLU, the derivative can also display an appropriate peak structure if the positive side is bounded due to the biological constraint of a saturated firing rate, making it a suitable balance function.

Consequently, the neuroplasticity update becomes:
\begin{equation}
    \Delta w_{ij} = \sum_{k,l} c_l(0) \frac{\partial \sigma(s_l)}{\partial s_l} w_{lk} \frac{\partial \sigma(s_k)}{\partial s_k} w_{ki} \frac{\partial \sigma(s_i)}{\partial s_i}x_j
\end{equation}
Since the neural firing dynamics have already converged to equilibrium by the time neuroplasticity occurs, the firing rates satisfy $\mathbf{x} = \sigma (\mathbf{W} \mathbf{x})$. Therefore, we have $\frac{\partial \sigma(s_l)}{\partial s_l} w_{lk} = \frac{\partial x_l}{\partial x_k}$ and $\frac{\partial \sigma(s_i)}{\partial s_i} x_j = \frac{\partial x_i}{\partial w_{ij}}$. The update then becomes:
\begin{equation}
    \Delta w_{ij} = \sum_{k,l} c_l(0) \frac{\partial x_l}{\partial x_k} \frac{\partial x_k}{\partial x_i} \frac{\partial x_i}{\partial w_{ij}}
\end{equation}
In this form, we intuitively demonstrate how the credit information in the $l^{\mathrm{th}}$ neuron can propagate to the $k^{\mathrm{th}}$ neuron, then to the $i^{\mathrm{th}}$ neuron, and finally reach the synapse $w_{ij}$. Although this derivation is informal, it reveals a connection between the timesteps of credit redistribution and the steps of the chain rule in BP. Therefore, by clamping the credit values of certain neurons with loss-related information, we can propagate the error signal through the neural network similarly to BP.

To formalize the connection we've outlined, we present a theorem that proves the equivalence of BP and our learning framework in a layered neural network (Figure \ref{fig:timescale}C):
\begin{theorem*}
Consider an $m$-layer neural network with neurons $\{\mathbf{x}^1, \mathbf{x}^2, \ldots, \mathbf{x}^m\}$ and connectivities $\{\mathbf{W}^1, \mathbf{W}^2, \ldots, \mathbf{W}^{m-1}\}$, governed by $\mathbf{x}^i = \sigma(\mathbf{W}^{i-1} \mathbf{x}^{i-1})$. Let the balance function be defined as $f = \partial \sigma / \partial s$. If the input layer's firing activity is clamped at $\mathbf{x}^1$ and the output layer's credit distribution is clamped at $-\partial L / \partial \mathbf{x}^m$, then $\Delta \mathbf{W} = -\partial L / \partial \mathbf{W}$.
\label{thm:example}
\end{theorem*} 

(For the detailed proof, please refer to the appendix.) This theorem indicates that simultaneous clamping of the input and output layers in a layered neural network can facilitate gradient descent on a loss function $L$ that is equivalent to BP. At first glance, the output clamping signal, $-\partial L / \partial \mathbf{x}^m$, seems to necessitate explicit derivative computation by neurons. However, this can often be simplified to algebraic computation, depending on the form of the loss function. For example, with the squared loss $L = \frac{1}{2} \|\mathbf{x}^m - \mathbf{y}\|^2$, the gradient $-\partial L / \partial \mathbf{x}^m$ simplifies to $\mathbf{y} - \mathbf{x}^m$. Therefore, the output clamping signal is merely the difference between the target activity $\mathbf{y}$ and the current activity $\mathbf{x}^m$. This difference computation is biologically plausible, as it can be implemented by other neural circuits.

\section{Simulation results: Learning rates dictate connectome patterns}
Next, we present simulations of our proposed framework. Given that the computational power of BP, to which our model reduces exactly, has been validated extensively, our simulation focuses not on the performance of our model on task datasets but on its predictions about neural connectomes. We investigate the classes of connectivity structures that can emerge from our neuroplasticity rule.

We note four key points about the set up of our simulation algorithm. \textit{First}, for the purposes of this exploration, the simulation algorithm makes one simplifying assumption 
In the model dynamics, the input/output clamping signal is replaced with a random Gaussian noise term, $\mathbf{\epsilon}$. This substitution serves the dual purposes of eliminating the need for specific input-loss pairs from open task datasets, and preventing the system from being trapped indefinitely in a trivial state. \textit{Second}, to ensure the existence of a non-trivial equilibrium states, the first 10\% of the rows of the matrix $\mathbf{J}$ were fixed to be a diagonal matrix. This can be interpreted as output clamping for that initial 10\% of neurons, thereby guaranteeing the existence of a non-trivial equilibrium state via clamping. \textit{Third}, in the wiring of biological networks, pruning of weak synaptic connections, a mechanism of structural plasticity, co-occurs with classical mechanisms of synaptic plasticity. Therefore, for the purposes of our exploration of neural connectivity structures, we pair our new synaptic plasticity rule with pruning. This encourages sparsity and the emergence of structured patterns of connectivity. \textit{Finally,} a learning rate factor, $r$, is added to Equation \ref{eq:dW rule}, which serves as a key free parameter of interest.

The pseudocode for our simulation is as follows:
\begin{algorithm}
\caption{Weight Update and Pruning}
\begin{algorithmic}[1] 
\State Initialize weight matrix $\mathbf{W}$, firing activity vector $\mathbf{x} \in \mathbb{R}^n$, credit vector $\mathbf{c} \in \mathbb{R}^n$
\State Set learning rate $r$, pruning probability $p$
\State Define activation function $\sigma$, balance function $f$, noise distribution $\mathbf{\epsilon}$
\For{$i = 1$ to $\text{max\_iteration}$}
    \State $\mathbf{x} \gets \sigma(\mathbf{Wx}) + \mathbf{\epsilon}$ \Comment{Update firing activity vector $\mathbf{x}$ with noise}
    \State $\mathbf{c} \gets \mathbf{Jc} + \mathbf{\epsilon}$ \Comment{Update credit vector $\mathbf{c}$ with noise}
    \State $\mathbf{W} \gets \mathbf{W} + r \cdot (\mathbf{c} \odot f(\mathbf{s})) \mathbf{x}^\top$ \Comment{Update weight matrix $\mathbf{W}$}
    \State Prune $\mathbf{W}$ by setting the $p$ proportion of smallest $|w_{ij}|$ to zero \Comment{Encourage sparsity}
\EndFor
\end{algorithmic}
\end{algorithm}

\begin{figure}[h!]
  \centering
  \includegraphics[width=0.99\textwidth]{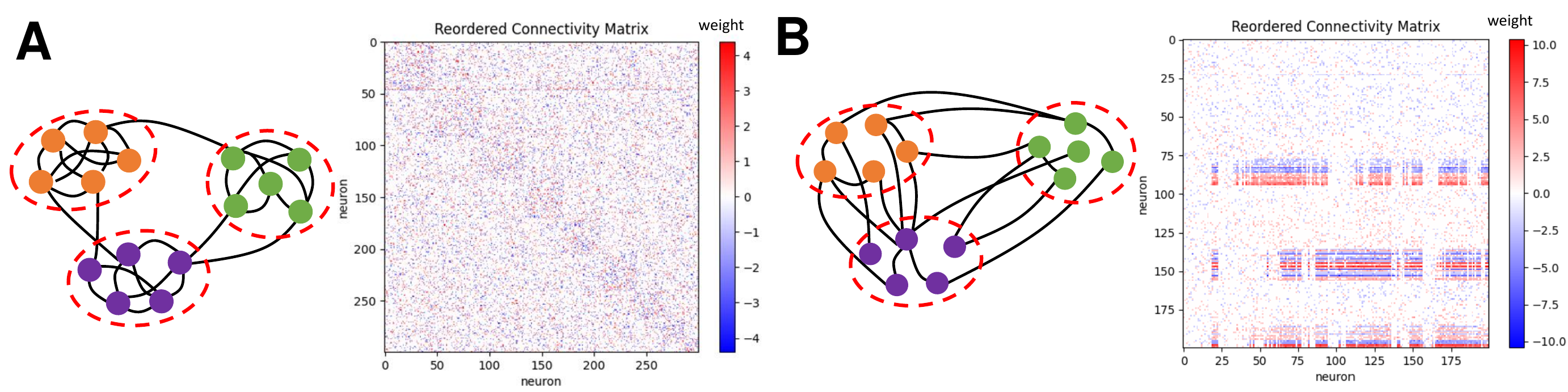}
  \caption{Simulation. Network connectivity structures with low (A) versus  high (B) learning rates.}
  \label{fig:simu}
  \vspace{-2em}
\end{figure}

We simulated an all-to-all connected neural network with 300 neurons. After running our simulation, we employed the Leiden algorithm \cite{traag2019louvain} to reorder the neurons in the search of community structure within this network. We discovered that with sufficiently strong pruning, different learning rates led to the emergence of distinct connectivity patterns in the network. At a low learning rate, a block-diagonal-like structure emerged (Figure \ref{fig:simu}A), indicating stronger within-module connectivity. Conversely, at a high learning rate, we observed prominent off-diagonal block connectivity (Figure \ref{fig:simu}B), suggesting stronger across-module connectivity. The computational requirements for this simulation are lightweight, allowing it to be run on a personal computer within minutes (Configuration used for simulation: Intel Core i7-14700F CPU, 32GB RAM, NVIDIA RTX 4070 GPU).

\section{Biological plausibility of our neuroplasticity framework}
Here, we address the biological correspondence of the key aspects of our framework.

\textbf{Credit and Neurotrophic factors.} We introduced a novel factor in our model, namely, the credit value $c$. Just as neural firing rates are taken to encode $x$, it is important to clarify what neural feature may encode $c$, or possibly the product of the credit and balance value, $cf$. We propose that neurotrophic factors (NTFs) could satisfy this role. NTFs, which are retrograde messengers, can be involved in retrograde credit redistribution processes. Moreover, NTFs are critical for neuronal survival, growth, maintenance, and repair \cite{davies1996neurotrophic, yuen1996nerve, mendell1999neurotrophin}, for neural plasticity \cite{thoenen1995neurotrophins, mcallister1999neurotrophins}, and for mediating E-I balance \cite{rutherford1997brain, desai1999bdnf}. NTFs are, therefore, plausible candidates to encode $c$ or $cf$.

\textbf{Backward credit assignment and Retrograde Signaling.}
Retrograde signaling involves the release of a retrograde messenger by a postsynaptic dendrite or cell body, which then travels backward across a chemical synapse to bind to the axon terminal of a presynaptic neuron \cite{tao2001retrograde}. Different types of retrograde messengers have been identified experimentally: gaseous types such as nitric oxide (NO) \cite{o1991tests, malen1997nitric}, endocannabinoids such as anandamide \cite{alger2002retrograde, wilson2001endogenous, kreitzer2002retrograde}, neurotrophic factors such as neurotrophin \cite{poo2001neurotrophins, ginty2002retrograde} and nerve growth factor \cite{yuen1996nerve}, prostaglandins \cite{sang2005postsynaptically, li2018prostaglandin}, and classical neurotransmitters \cite{duguid2004retrograde, regehr2009activity}. 
Although retrograde signaling systems can vary in the physical spread of their action (affecting synapses in an input-specific way versus affecting millions of synapses) and in the speed of their reuptake and degradation \cite{wilson2001endogenous, regehr2009activity, harward2016autocrine, wood1994models, philippides2000four}, their time-of-action is usually within seconds to minutes \cite{heifets2009endocannabinoid, park2013neurotrophin, ahern2002cgmp}, consistent with model requirements.


\textbf{Multiple timescales of operation and Neural oscillations.}
The brain operates on hierarchical timescales with nested dynamics \cite{kaplan2020nested, wolff2022intrinsic}. Neural oscillations, the rhythmic patterns of activity in the nervous system, reflect this hierarchy by spanning a wide range of frequencies. In the mammalian forebrain, for example, networks exhibit oscillatory bands from 0.05 Hz to 500 Hz \cite{buzsaki2004neuronal}. It has been suggested that neural oscillations are crucial for coordinating activity across timescales, providing phase information \cite{canolty2010functional, lisman2013theta}, and in aiding functional integration \cite{palva2018roles}. Neural oscillations are also involved in neuroplasticity \cite{buzsaki2002theta, steriade2003neuronal} and are closely related to E-I balance \cite{atallah2009instantaneous, isaacson2011inhibition}.

\textbf{Input/output clamping and Associative learning.} Our model requires clamping some input neurons with input signals and output neurons with loss-signal information. For the network to update weights, the input clamping period must overlap sufficiently with the output clamping period to allow plasticity. The longer the overlap, the more updates occur, resembling the temporal contiguity feature in associative learning for both classical and operant conditioning. The notion of input/output clamping is, therefore, congruent with typical scenarios of animals interacting with their environments.

\section{Discussion}

In this work, we proposed a simple neuroplasticity rule rooted in the  principle of E-I balance, and with a functional form reminiscent of the classical Hebbian rule. Strikingly, we show that in layered neural networks, it is isomorphic to 
the BP algorithm under the sole assumption that the functional metric quantifying the E-I balance of a neuron be chosen to have the functional form of the derivative of its activation function. Moreover, the model is biologically plausible under two further assumptions: hierarchical timescales, and retrograde signaling, both of which have well-established biological mechanisms in neuroscience. In the framework of this model, BP can be interpreted as a simple consequence of individual neurons striving to maintain E-I balance. Our simulations demonstrate that this synaptic plasticity rule, when paired with structural plasticity mechanisms like pruning, can lead to the self-organization of different network structures in a fully connected network.


Our model makes three practical predictions that are experimentally testable. \textbf{\textit{First}}, the amplitude of retrograde signaling should be highest for balanced neurons, but low for neurons with saturated firing rates or those close to being silent. This is based on the equation $g_{ji}=c_i f(s_i) w_{ij}$, where the balance value $f$ approaches zero as a neuron becomes more unbalanced. \textbf{\textit{Second}}, the ratio of the amplitude of synaptic change to the amplitude of retrograde signal should be proportional to the firing rate of the presynaptic neuron and inversely proportional to synaptic strength. This prediction is derived from the ratio of the credit redistribution rule to the neuroplasticity rule, leading to $\frac{\Delta w_{ij}}{g_{ji}} = \frac{x_j}{w_{ij}}$. \textbf{\textit{Third}}, during early stages of neural development, associated with a greater degree of plasticity and therefore higher learning rate, the brain is more likely to form across-module connectivity. By contrast, in later stages, associated with lower learning rates, within-module connectivity is more likely. 

Our model also reveals an intriguing link between BP and associative learning. First, the core formulation of our learning rule (that reduces exactly to BP) is Hebbian-like, promoting parallels with associative learning. Second, the  clamping of input and loss onto the "input" and "output" neurons for overlapping durations are reminiscent of the temporal contiguity needed in associative learning paradigms. In this sense, BP can be viewed as associative learning in a broader context. 

The model we have proposed here possesses a potential limitation. It is the assumption that credit redistribution will converge for the network updating rule to function. Our current mathematical framework does not guarantee the existence of a non-trivial fixed point in the dynamical system for credit redistribution. The matrix $\mathbf{J}$ in Equation 5 does not necessarily have an eigenvalue of one, meaning it could result in convergence to a zero vector or divergence to infinity. We currently circumvent this issue through the plausible means of clamping the credit values of some neurons (technically, even of just one neuron) to a constant. With such clamping, multiple ($\ge 1$) rows of the matrix $\mathbf{J}$ will be replaced by an identity matrix, ensuring the existence of an eigenvalue of one. Consequently, an intriguing observation is that the network's stability may depend on external credit signals and that spontaneous credit dynamics could lead to instability. Long-term sensory deprivation, which causes cognitive instability \cite{bexton1954effects, solomon1957sensory} and affects connectivity and E-I balance in cortical circuits \cite{maffei2009network}, is one biological phenomenon that supports this observation. More generally, then, the stability of this model will benefit from further exploration and systematic analysis.

\section*{Acknowledgements}
This work was supported by funding from a JHU Discovery Award co-funded with the One Neuro Initiative (SPM), and from NIH grant 2R01EY027718 (SPM). We thank Hokin Deng (JHU) for making helpful suggestions about the title of the manuscript.

\bibliographystyle{unsrt}
\bibliography{refs_nips2024}

\begin{thebibliography}{10}

\bibitem{krizhevsky2012imagenet}
Alex Krizhevsky, Ilya Sutskever, and Geoffrey~E Hinton.
\newblock Imagenet classification with deep convolutional neural networks.
\newblock {\em Advances in neural information processing systems}, 25, 2012.

\bibitem{amodei2016deep}
Dario Amodei, Sundaram Ananthanarayanan, Rishita Anubhai, Jingliang Bai, Eric Battenberg, Carl Case, Jared Casper, Bryan Catanzaro, Qiang Cheng, Guoliang Chen, et~al.
\newblock Deep speech 2: End-to-end speech recognition in english and mandarin.
\newblock In {\em International conference on machine learning}, pages 173--182. PMLR, 2016.

\bibitem{vaswani2017attention}
Ashish Vaswani, Noam Shazeer, Niki Parmar, Jakob Uszkoreit, Llion Jones, Aidan~N Gomez, {\L}ukasz Kaiser, and Illia Polosukhin.
\newblock Attention is all you need.
\newblock {\em Advances in neural information processing systems}, 30, 2017.

\bibitem{sezer2020financial}
Omer~Berat Sezer, Mehmet~Ugur Gudelek, and Ahmet~Murat Ozbayoglu.
\newblock Financial time series forecasting with deep learning: A systematic literature review: 2005--2019.
\newblock {\em Applied soft computing}, 90:106181, 2020.

\bibitem{rajpurkar2017chexnet}
Pranav Rajpurkar, Jeremy Irvin, Kaylie Zhu, Brandon Yang, Hershel Mehta, Tony Duan, Daisy Ding, Aarti Bagul, Curtis Langlotz, Katie Shpanskaya, et~al.
\newblock Chexnet: Radiologist-level pneumonia detection on chest x-rays with deep learning.
\newblock {\em arXiv preprint arXiv:1711.05225}, 2017.

\bibitem{lillicrap2016random}
Timothy~P Lillicrap, Daniel Cownden, Douglas~B Tweed, and Colin~J Akerman.
\newblock Random synaptic feedback weights support error backpropagation for deep learning.
\newblock {\em Nature communications}, 7(1):13276, 2016.

\bibitem{liao2016important}
Qianli Liao, Joel Leibo, and Tomaso Poggio.
\newblock How important is weight symmetry in backpropagation?
\newblock In {\em Proceedings of the AAAI Conference on Artificial Intelligence}, volume~30, 2016.

\bibitem{russakovsky2015imagenet}
Olga Russakovsky, Jia Deng, Hao Su, Jonathan Krause, Sanjeev Satheesh, Sean Ma, Zhiheng Huang, Andrej Karpathy, Aditya Khosla, Michael Bernstein, et~al.
\newblock Imagenet large scale visual recognition challenge.
\newblock {\em International journal of computer vision}, 115:211--252, 2015.

\bibitem{bartunov2018assessing}
Sergey Bartunov, Adam Santoro, Blake Richards, Luke Marris, Geoffrey~E Hinton, and Timothy Lillicrap.
\newblock Assessing the scalability of biologically-motivated deep learning algorithms and architectures.
\newblock {\em Advances in neural information processing systems}, 31, 2018.

\bibitem{akrout2019deep}
Mohamed Akrout, Collin Wilson, Peter Humphreys, Timothy Lillicrap, and Douglas~B Tweed.
\newblock Deep learning without weight transport.
\newblock {\em Advances in neural information processing systems}, 32, 2019.

\bibitem{nokland2016direct}
Arild N{\o}kland.
\newblock Direct feedback alignment provides learning in deep neural networks.
\newblock {\em Advances in neural information processing systems}, 29, 2016.

\bibitem{clark2021credit}
David Clark, LF~Abbott, and SueYeon Chung.
\newblock Credit assignment through broadcasting a global error vector.
\newblock {\em Advances in Neural Information Processing Systems}, 34:10053--10066, 2021.

\bibitem{le1986learning}
Yann Le~Cun.
\newblock Learning process in an asymmetric threshold network.
\newblock In {\em Disordered systems and biological organization}, pages 233--240. Springer, 1986.

\bibitem{bengio2014auto}
Yoshua Bengio.
\newblock How auto-encoders could provide credit assignment in deep networks via target propagation.
\newblock {\em arXiv preprint arXiv:1407.7906}, 2014.

\bibitem{meulemans2020theoretical}
Alexander Meulemans, Francesco Carzaniga, Johan Suykens, Jo{\~a}o Sacramento, and Benjamin~F Grewe.
\newblock A theoretical framework for target propagation.
\newblock {\em Advances in Neural Information Processing Systems}, 33:20024--20036, 2020.

\bibitem{lee2015difference}
Dong-Hyun Lee, Saizheng Zhang, Asja Fischer, and Yoshua Bengio.
\newblock Difference target propagation.
\newblock In {\em Machine Learning and Knowledge Discovery in Databases: European Conference, ECML PKDD 2015, Porto, Portugal, September 7-11, 2015, Proceedings, Part I 15}, pages 498--515. Springer, 2015.

\bibitem{frenkel2021learning}
Charlotte Frenkel, Martin Lefebvre, and David Bol.
\newblock Learning without feedback: Fixed random learning signals allow for feedforward training of deep neural networks.
\newblock {\em Frontiers in neuroscience}, 15:629892, 2021.

\bibitem{lansdell2019learning}
Benjamin~James Lansdell, Prashanth~Ravi Prakash, and Konrad~Paul Kording.
\newblock Learning to solve the credit assignment problem.
\newblock {\em arXiv preprint arXiv:1906.00889}, 2019.

\bibitem{meulemans2021credit}
Alexander Meulemans, Matilde Tristany~Farinha, Javier Garc{\'\i}a~Ord{\'o}{\~n}ez, Pau Vilimelis~Aceituno, Jo{\~a}o Sacramento, and Benjamin~F Grewe.
\newblock Credit assignment in neural networks through deep feedback control.
\newblock {\em Advances in Neural Information Processing Systems}, 34:4674--4687, 2021.

\bibitem{meulemans2022least}
Alexander Meulemans, Nicolas Zucchet, Seijin Kobayashi, Johannes Von~Oswald, and Jo{\~a}o Sacramento.
\newblock The least-control principle for local learning at equilibrium.
\newblock {\em Advances in Neural Information Processing Systems}, 35:33603--33617, 2022.

\bibitem{harris2008stability}
Kenneth~D Harris.
\newblock Stability of the fittest: organizing learning through retroaxonal signals.
\newblock {\em Trends in neurosciences}, 31(3):130--136, 2008.

\bibitem{lillicrap2020backpropagation}
Timothy~P Lillicrap, Adam Santoro, Luke Marris, Colin~J Akerman, and Geoffrey Hinton.
\newblock Backpropagation and the brain.
\newblock {\em Nature Reviews Neuroscience}, 21(6):335--346, 2020.

\bibitem{oztas2003neuronal}
Emin Oztas et~al.
\newblock Neuronal tracing.
\newblock {\em Neuroanatomy}, 2(2):5, 2003.

\bibitem{friston2005theory}
Karl Friston.
\newblock A theory of cortical responses.
\newblock {\em Philosophical transactions of the Royal Society B: Biological sciences}, 360(1456):815--836, 2005.

\bibitem{whittington2017approximation}
James~CR Whittington and Rafal Bogacz.
\newblock An approximation of the error backpropagation algorithm in a predictive coding network with local hebbian synaptic plasticity.
\newblock {\em Neural computation}, 29(5):1229--1262, 2017.

\bibitem{scellier2017equilibrium}
Benjamin Scellier and Yoshua Bengio.
\newblock Equilibrium propagation: Bridging the gap between energy-based models and backpropagation.
\newblock {\em Frontiers in computational neuroscience}, 11:24, 2017.

\bibitem{scellier2019equivalence}
Benjamin Scellier and Yoshua Bengio.
\newblock Equivalence of equilibrium propagation and recurrent backpropagation.
\newblock {\em Neural computation}, 31(2):312--329, 2019.

\bibitem{laborieux2022holomorphic}
Axel Laborieux and Friedemann Zenke.
\newblock Holomorphic equilibrium propagation computes exact gradients through finite size oscillations.
\newblock {\em Advances in Neural Information Processing Systems}, 35:12950--12963, 2022.

\bibitem{payeur2021burst}
Alexandre Payeur, Jordan Guerguiev, Friedemann Zenke, Blake~A Richards, and Richard Naud.
\newblock Burst-dependent synaptic plasticity can coordinate learning in hierarchical circuits.
\newblock {\em Nature neuroscience}, 24(7):1010--1019, 2021.

\bibitem{movellan1991contrastive}
Javier~R Movellan.
\newblock Contrastive hebbian learning in the continuous hopfield model.
\newblock In {\em Connectionist models}, pages 10--17. Elsevier, 1991.

\bibitem{xie2003equivalence}
Xiaohui Xie and H~Sebastian Seung.
\newblock Equivalence of backpropagation and contrastive hebbian learning in a layered network.
\newblock {\em Neural computation}, 15(2):441--454, 2003.

\bibitem{hinton1987learning}
Geoffrey~E Hinton and James McClelland.
\newblock Learning representations by recirculation.
\newblock In {\em Neural information processing systems}, 1987.

\bibitem{o1996biologically}
Randall~C O'Reilly.
\newblock Biologically plausible error-driven learning using local activation differences: The generalized recirculation algorithm.
\newblock {\em Neural computation}, 8(5):895--938, 1996.

\bibitem{van1996chaos}
Carl Van~Vreeswijk and Haim Sompolinsky.
\newblock Chaos in neuronal networks with balanced excitatory and inhibitory activity.
\newblock {\em Science}, 274(5293):1724--1726, 1996.

\bibitem{deneve2016efficient}
Sophie Den{\`e}ve and Christian~K Machens.
\newblock Efficient codes and balanced networks.
\newblock {\em Nature neuroscience}, 19(3):375--382, 2016.

\bibitem{zhou2018synaptic}
Shanglin Zhou and Yuguo Yu.
\newblock Synaptic ei balance underlies efficient neural coding.
\newblock {\em Frontiers in neuroscience}, 12:307227, 2018.

\bibitem{turrigiano2004homeostatic}
Gina~G Turrigiano and Sacha~B Nelson.
\newblock Homeostatic plasticity in the developing nervous system.
\newblock {\em Nature reviews neuroscience}, 5(2):97--107, 2004.

\bibitem{turrigiano2008self}
Gina~G Turrigiano.
\newblock The self-tuning neuron: synaptic scaling of excitatory synapses.
\newblock {\em Cell}, 135(3):422--435, 2008.

\bibitem{desai1999plasticity}
Niraj~S Desai, Lana~C Rutherford, and Gina~G Turrigiano.
\newblock Plasticity in the intrinsic excitability of cortical pyramidal neurons.
\newblock {\em Nature neuroscience}, 2(6):515--520, 1999.

\bibitem{field2020heterosynaptic}
Rachel~E Field, James~A D¡¯amour, Robin Tremblay, Christoph Miehl, Bernardo Rudy, Julijana Gjorgjieva, and Robert~C Froemke.
\newblock Heterosynaptic plasticity determines the set point for cortical excitatory-inhibitory balance.
\newblock {\em Neuron}, 106(5):842--854, 2020.

\bibitem{traag2019louvain}
Vincent~A Traag, Ludo Waltman, and Nees~Jan Van~Eck.
\newblock From louvain to leiden: guaranteeing well-connected communities.
\newblock {\em Scientific reports}, 9(1):5233, 2019.

\bibitem{davies1996neurotrophic}
Alun~Milward Davies.
\newblock The neurotrophic hypothesis: where does it stand?
\newblock {\em Philosophical Transactions of the Royal Society of London. Series B: Biological Sciences}, 351(1338):389--394, 1996.

\bibitem{yuen1996nerve}
Eric~C Yuen, Charles~L Howe, Yiwen Li, David~M Holtzman, and William~C Mobley.
\newblock Nerve growth factor and the neurotrophic factor hypothesis.
\newblock {\em Brain and Development}, 18(5):362--368, 1996.

\bibitem{mendell1999neurotrophin}
Lorne~M Mendell.
\newblock Neurotrophin action on sensory neurons in adults: an extension of the neurotrophic hypothesis.
\newblock {\em Pain}, 82:S127--S132, 1999.

\bibitem{thoenen1995neurotrophins}
Hans Thoenen.
\newblock Neurotrophins and neuronal plasticity.
\newblock {\em Science}, 270(5236):593--598, 1995.

\bibitem{mcallister1999neurotrophins}
A~Kimberley McAllister, Lawrence~C Katz, and Donald~C Lo.
\newblock Neurotrophins and synaptic plasticity.
\newblock {\em Annual review of neuroscience}, 22(1):295--318, 1999.

\bibitem{rutherford1997brain}
Lana~C Rutherford, Andrew DeWan, Holly~M Lauer, and Gina~G Turrigiano.
\newblock Brain-derived neurotrophic factor mediates the activity-dependent regulation of inhibition in neocortical cultures.
\newblock {\em Journal of Neuroscience}, 17(12):4527--4535, 1997.

\bibitem{desai1999bdnf}
Niraj~S Desai, Lana~C Rutherford, and Gina~G Turrigiano.
\newblock Bdnf regulates the intrinsic excitability of cortical neurons.
\newblock {\em Learning \& Memory}, 6(3):284--291, 1999.

\bibitem{tao2001retrograde}
Huizhong~W Tao and Mu-ming Poo.
\newblock Retrograde signaling at central synapses.
\newblock {\em Proceedings of the National Academy of Sciences}, 98(20):11009--11015, 2001.

\bibitem{o1991tests}
Thomas~J O'dell, Robert~D Hawkins, Eric~R Kandel, and Ottavio Arancio.
\newblock Tests of the roles of two diffusible substances in long-term potentiation: evidence for nitric oxide as a possible early retrograde messenger.
\newblock {\em Proceedings of the National Academy of Sciences}, 88(24):11285--11289, 1991.

\bibitem{malen1997nitric}
Peter~L Malen and Paul~F Chapman.
\newblock Nitric oxide facilitates long-term potentiation, but not long-term depression.
\newblock {\em Journal of Neuroscience}, 17(7):2645--2651, 1997.

\bibitem{alger2002retrograde}
Bradley~E Alger.
\newblock Retrograde signaling in the regulation of synaptic transmission: focus on endocannabinoids.
\newblock {\em Progress in neurobiology}, 68(4):247--286, 2002.

\bibitem{wilson2001endogenous}
Rachel~I Wilson and Roger~A Nicoll.
\newblock Endogenous cannabinoids mediate retrograde signalling at hippocampal synapses.
\newblock {\em Nature}, 410(6828):588--592, 2001.

\bibitem{kreitzer2002retrograde}
Anatol~C Kreitzer and Wade~G Regehr.
\newblock Retrograde signaling by endocannabinoids.
\newblock {\em Current opinion in neurobiology}, 12(3):324--330, 2002.

\bibitem{poo2001neurotrophins}
Mu-ming Poo.
\newblock Neurotrophins as synaptic modulators.
\newblock {\em Nature reviews neuroscience}, 2(1):24--32, 2001.

\bibitem{ginty2002retrograde}
David~D Ginty and Rosalind~A Segal.
\newblock Retrograde neurotrophin signaling: Trk-ing along the axon.
\newblock {\em Current opinion in neurobiology}, 12(3):268--274, 2002.

\bibitem{sang2005postsynaptically}
Nan Sang, Jian Zhang, Victor Marcheselli, Nicolas~G Bazan, and Chu Chen.
\newblock Postsynaptically synthesized prostaglandin e2 (pge2) modulates hippocampal synaptic transmission via a presynaptic pge2 ep2 receptor.
\newblock {\em Journal of Neuroscience}, 25(43):9858--9870, 2005.

\bibitem{li2018prostaglandin}
Jie Li, Elizabeth Serafin, and Mark~L Baccei.
\newblock Prostaglandin signaling governs spike timing-dependent plasticity at sensory synapses onto mouse spinal projection neurons.
\newblock {\em Journal of Neuroscience}, 38(30):6628--6639, 2018.

\bibitem{duguid2004retrograde}
Ian~C Duguid and Trevor~G Smart.
\newblock Retrograde activation of presynaptic nmda receptors enhances gaba release at cerebellar interneuron--purkinje cell synapses.
\newblock {\em Nature neuroscience}, 7(5):525--533, 2004.

\bibitem{regehr2009activity}
Wade~G Regehr, Megan~R Carey, and Aaron~R Best.
\newblock Activity-dependent regulation of synapses by retrograde messengers.
\newblock {\em Neuron}, 63(2):154--170, 2009.

\bibitem{harward2016autocrine}
Stephen~C Harward, Nathan~G Hedrick, Charles~E Hall, Paula Parra-Bueno, Teresa~A Milner, Enhui Pan, Tal Laviv, Barbara~L Hempstead, Ryohei Yasuda, and James~O McNamara.
\newblock Autocrine bdnf--trkb signalling within a single dendritic spine.
\newblock {\em Nature}, 538(7623):99--103, 2016.

\bibitem{wood1994models}
J~Wood and J~Garthwaite.
\newblock Models of the diffusional spread of nitric oxide: implications for neural nitric oxide signalling and its pharmacological properties.
\newblock {\em Neuropharmacology}, 33(11):1235--1244, 1994.

\bibitem{philippides2000four}
Andrew Philippides, Phil Husbands, and Michael O'Shea.
\newblock Four-dimensional neuronal signaling by nitric oxide: a computational analysis.
\newblock {\em Journal of Neuroscience}, 20(3):1199--1207, 2000.

\bibitem{heifets2009endocannabinoid}
Boris~D Heifets and Pablo~E Castillo.
\newblock Endocannabinoid signaling and long-term synaptic plasticity.
\newblock {\em Annual review of physiology}, 71:283--306, 2009.

\bibitem{park2013neurotrophin}
Hyungju Park and Mu-ming Poo.
\newblock Neurotrophin regulation of neural circuit development and function.
\newblock {\em Nature Reviews Neuroscience}, 14(1):7--23, 2013.

\bibitem{ahern2002cgmp}
Gerard~P Ahern, Vitaly~A Klyachko, and Meyer~B Jackson.
\newblock cgmp and s-nitrosylation: two routes for modulation of neuronal excitability by no.
\newblock {\em Trends in neurosciences}, 25(10):510--517, 2002.

\bibitem{kaplan2020nested}
Harris~S Kaplan, Oriana~Salazar Thula, Niklas Khoss, and Manuel Zimmer.
\newblock Nested neuronal dynamics orchestrate a behavioral hierarchy across timescales.
\newblock {\em Neuron}, 105(3):562--576, 2020.

\bibitem{wolff2022intrinsic}
Annemarie Wolff, Nareg Berberian, Mehrshad Golesorkhi, Javier Gomez-Pilar, Federico Zilio, and Georg Northoff.
\newblock Intrinsic neural timescales: temporal integration and segregation.
\newblock {\em Trends in cognitive sciences}, 26(2):159--173, 2022.

\bibitem{buzsaki2004neuronal}
Gyorgy Buzsaki and Andreas Draguhn.
\newblock Neuronal oscillations in cortical networks.
\newblock {\em science}, 304(5679):1926--1929, 2004.

\bibitem{canolty2010functional}
Ryan~T Canolty and Robert~T Knight.
\newblock The functional role of cross-frequency coupling.
\newblock {\em Trends in cognitive sciences}, 14(11):506--515, 2010.

\bibitem{lisman2013theta}
John~E Lisman and Ole Jensen.
\newblock The theta-gamma neural code.
\newblock {\em Neuron}, 77(6):1002--1016, 2013.

\bibitem{palva2018roles}
Satu Palva and J~Matias Palva.
\newblock Roles of brain criticality and multiscale oscillations in temporal predictions for sensorimotor processing.
\newblock {\em Trends in neurosciences}, 41(10):729--743, 2018.

\bibitem{buzsaki2002theta}
Gy{\"o}rgy Buzs{\'a}ki.
\newblock Theta oscillations in the hippocampus.
\newblock {\em Neuron}, 33(3):325--340, 2002.

\bibitem{steriade2003neuronal}
Mircea Steriade and Igor Timofeev.
\newblock Neuronal plasticity in thalamocortical networks during sleep and waking oscillations.
\newblock {\em Neuron}, 37(4):563--576, 2003.

\bibitem{atallah2009instantaneous}
Bassam~V Atallah and Massimo Scanziani.
\newblock Instantaneous modulation of gamma oscillation frequency by balancing excitation with inhibition.
\newblock {\em Neuron}, 62(4):566--577, 2009.

\bibitem{isaacson2011inhibition}
Jeffry~S Isaacson and Massimo Scanziani.
\newblock How inhibition shapes cortical activity.
\newblock {\em Neuron}, 72(2):231--243, 2011.

\bibitem{bexton1954effects}
William~Harold Bexton, Woodburn Heron, and Thomas~H Scott.
\newblock Effects of decreased variation in the sensory environment.
\newblock {\em Canadian Journal of Psychology/Revue canadienne de psychologie}, 8(2):70, 1954.

\bibitem{solomon1957sensory}
Philip Solomon, P~Herbert Leiderman, Jack Mendelson, and Donald Wexler.
\newblock Sensory deprivation: A review.
\newblock {\em American Journal of Psychiatry}, 114(4):357--363, 1957.

\bibitem{maffei2009network}
Arianna Maffei and Alfredo Fontanini.
\newblock Network homeostasis: a matter of coordination.
\newblock {\em Current opinion in neurobiology}, 19(2):168--173, 2009.

\end{thebibliography}

\newpage
\section{Appendix}

\subsection{Link to Backpropagation: Proof for layered neural network}
\begin{theorem*}
Consider an $m$-layer neural network with neurons $\{\mathbf{x}^1, \mathbf{x}^2, \ldots, \mathbf{x}^m\}$ and connectivities $\{\mathbf{W}^1, \mathbf{W}^2, \ldots, \mathbf{W}^{m-1}\}$, governed by $\mathbf{x}^i = \sigma(\mathbf{W}^{i-1} \mathbf{x}^{i-1})$. Let the balance function be defined as $f = \partial \sigma / \partial s$. If the input layer's firing activity is clamped at $\mathbf{x}^1$ and the output layer's credit distribution is clamped at $-\partial L / \partial \mathbf{x}^m$, then $\Delta \mathbf{W} = -\partial L / \partial \mathbf{W}$.
\label{thm:example}
\end{theorem*}
\textit{Proof.} 
We consider a typical layered neural network: the network has $m$ layers, with $n_i$ neurons in the $i^{\text{th}}$ layer, and their activities are represented as a column vector $\mathbf{x}^i = (x^i_1, x^i_2, \ldots, x^i_{n_i})$. The overall activity of the neural network is the concatenation of all layers, $\mathbf{x} = (\mathbf{x}^1, \mathbf{x}^2, \ldots, \mathbf{x}^m)$. Each layer $\mathbf{x}^i$ receives input solely from the previous layer $\mathbf{x}^{i-1}$ through the connectivity matrix $\mathbf{W}^{i-1}$. Consequently, the firing dynamics of this layered network can be described as:
\begin{equation}
\mathbf{x}(t''+1) = \sigma(\mathbf{W}\mathbf{x}(t'')) = \sigma(
\left[
\setlength\arraycolsep{2pt}
\begin{array}{ccccc}
\mathbf{0}      &              &              &                  &   \\ [0.8ex]
\mathbf{W}^1    & \mathbf{0}   &              &                  &  \\ [1.2ex]
                & \mathbf{W}^2 & \mathbf{0}   &                  &  \\ [0.5ex]
                &              & \ddots       & \ddots           &  \\ [1.ex]
                &              &              & \mathbf{W}^{m-1} & \mathbf{0}
\end{array}
\right]
\begin{bmatrix}
\mathbf{x}^1(t'') \\
\mathbf{x}^2(t'') \\
\mathbf{x}^3(t'') \\
\vdots \\
\mathbf{x}^{m-1}(t'') \\
\mathbf{x}^m(t'') \\
\end{bmatrix}
)
\end{equation}
For this network, the \textit{neural activity} at equilibrium can be obtained by fixing the value of the input layer, i.e., keeping $\mathbf{x}^1(t'') \equiv \mathbf{x}^1$. For ease of explanation, if the network is initialized with $\mathbf{x} = (\mathbf{x}^1, \mathbf{0}, \mathbf{0}, \ldots)$, and $\mathbf{x}^1$ is kept fixed, reflecting a consistent external input signal, the firing activities will evolve as follows:
\begin{equation}
\begin{bmatrix}
\mathbf{x}^1 \\
\mathbf{0} \\
\mathbf{0} \\
\vdots \\
\mathbf{0} \\
\end{bmatrix} 
\rightarrow
\begin{bmatrix}
\mathbf{x}^1 \\
\mathbf{x}^2 = \sigma(\mathbf{W}^1\mathbf{x}^1)\\
\mathbf{0} \\
\vdots \\
\mathbf{0} \\
\end{bmatrix} 
\rightarrow
\begin{bmatrix}
\mathbf{x}^1 \\
\mathbf{x}^2 \\
\mathbf{x}^3 = \sigma(\mathbf{W}^2\mathbf{x}^2) \\
\vdots \\
\mathbf{0} \\
\end{bmatrix} 
\rightarrow
\begin{bmatrix}
\mathbf{x}^1 \\
\mathbf{x}^2 \\
\mathbf{x}^3 \\
\vdots \\
\mathbf{x}^m = \sigma(\mathbf{W}^{m-1}\mathbf{x}^{m-1}) \\
\end{bmatrix} 
\end{equation}
It can be observed that an $m$-layered neural network takes $m-1$ timesteps to reach equilibrium. Non-zero initialization in other layers does not affect the stable state because activities in successive postsynaptic layers will emerge as a result of inputs from their presynaptic layers.

Next, for \textit{credit distribution} in this $m$-layered neural network, we denote the credit values for the $i^{\text{th}}$ layer as $\mathbf{c}^i$. The entire network's credit distribution can thus be represented as a concatenated column vector $\mathbf{c} = (\mathbf{c}^1, \mathbf{c}^2, \ldots, \mathbf{c}^m)$. If the balance function is chosen as $f = \frac{\partial \sigma(s)}{\partial s}$, the elements in the $\mathbf{J}$ matrix become derivatives between pairs of neurons. Representing this in block matrix form:
\begin{equation}
\mathbf{c}(t'+1) = \mathbf{J}\mathbf{c}(t') = \left[
\setlength\arraycolsep{1pt}
\begin{array}{cccccc}
\mathbf{0}  & \; (\dfrac{\partial \mathbf{x}^2}{ \partial \mathbf{x}^1})^T    &  &   &   &\\ [0.8ex]
& \mathbf{0}&(\dfrac{\partial \mathbf{x}^3}{ \partial \mathbf{x}^2})^T&&&  \\ [1.2ex]
&&&&&\\ [-1.5ex]
&&\ddots& \ddots&&\\ [0.ex]
&&&\mathbf{0}&(\dfrac{\partial \mathbf{x}^m}{ \partial \mathbf{x}^{m-1}})^T&\\ [2.0ex]
&&&& \mathbf{0}
\end{array}
\right]
\left[
\begin{array}{c}
\mathbf{c}^1(t')\\ [1pt]
\mathbf{c}^2(t') \\ [3pt]
\mathbf{c}^3(t') \\ [6pt]
\vdots \\ [8pt]
\mathbf{c}^{m-1}(t') \\ [4pt]
\mathbf{c}^m(t') \\ 
\end{array}
\right]
\end{equation}
where the blocks are transposes of Jacobian matrices. For example,
\begin{equation}
(\frac{\partial \mathbf{x}^m}{ \partial \mathbf{x}^{m-1}})^T = 
\left[
\setlength\arraycolsep{3pt}
\begin{array}{cccc}
\dfrac{\partial x^m_1}{ \partial x^{m-1}_1}  & \dfrac{\partial x^m_2}{ \partial x^{m-1}_1}     & \hdots  &  \dfrac{\partial x^m_{n_m}}{ \partial x^{m-1}_1}\\ [12pt]
\dfrac{\partial x^m_1}{ \partial x^{m-1}_2}& \dfrac{\partial x^m_2}{ \partial x^{m-1}_2}&\hdots& \dfrac{\partial x^m_{n_m}}{ \partial x^{m-1}_2} \\ [5pt]
\vdots&\vdots&\ddots&\vdots\\ [5pt]
\dfrac{\partial x^m_1}{ \partial x^{m-1}_{n_{m-1}}}&\dfrac{\partial x^m_2}{ \partial x^{m-1}_{n_{m-1}}}&\hdots&\dfrac{\partial x^m_{n_m}}{ \partial x^{m-1}_{n_{m-1}}}\\ 
\end{array}
\right]
\end{equation}
The stable state of the credit distribution can be obtained by fixing the credits at the output layer, reflecting  
a consistent external signal:  $\mathbf{c}^m(t) \equiv \mathbf{c}^m$. For simplicity, if the other layers are initialized with zero credits, the credit distribution will evolve as follows:
\begin{equation}
\renewcommand{\arraystretch}{1.2}
\begin{bmatrix}
\mathbf{0} \\
\vdots \\
\mathbf{0} \\
\mathbf{0} \\
\mathbf{c}^m \\
\end{bmatrix} 
\rightarrow
\begin{bmatrix}
\mathbf{0} \\
\vdots \\
\mathbf{0} \\
\mathbf{c}^{m-1} = (\dfrac{\partial \mathbf{x}^m}{ \partial \mathbf{x}^{m-1}})^T \mathbf{c}^m \\
\mathbf{c}^m \\
\end{bmatrix} 
\rightarrow
\begin{bmatrix}
\mathbf{0} \\
\vdots \\
\mathbf{c}^{m-2} = (\dfrac{\partial \mathbf{x}^{m-1}}{ \partial \mathbf{x}^{m-2}})^T \mathbf{c}^{m-1} \\
\mathbf{c}^{m-1}\\
\mathbf{c}^m \\
\end{bmatrix} 
\rightarrow
\begin{bmatrix}
\mathbf{c}^1 = (\dfrac{\partial \mathbf{x}^2}{ \partial \mathbf{x}^1})^T \mathbf{c}^2 \\
\vdots \\
\mathbf{c}^{m-2}\\
\mathbf{c}^{m-1}\\
\mathbf{c}^m \\
\end{bmatrix} 
\end{equation}
Therefore, it takes $m-1$ timesteps for an $m$-layered network to reach equilibrium. In this scenario, the converged distribution depends solely on the consistent activity of the output layer. Initializing other layers with non-zero values will not affect the stable distribution.

Finally, we consider \textit{updates to network weights} by the application of our neuroplasticity rule. Since we have specified the balance function $f$ as $\partial \sigma / \partial s$, the original rule becomes:
\begin{equation}
    \Delta w_{ij} = c_i \sigma'(s_i) x_j
\end{equation}
Additionally, given the layered network structure, updates for all illegal connections (cross-layer, within-layer, or recurrent connections) are disregarded, and these connections remain unconnected. The update for connections between the $k^{\text{th}}$ and $(k+1)^{\text{th}}$ layers is given by:
\begin{equation}
    \Delta \mathbf{W}^k = \sigma'( \mathbf{s}^{k+1}) \odot \mathbf{c}^{k+1}   \; (\mathbf{x}^k)^T
\end{equation}
where $\mathbf{s}^{k+1}$ is the vector summarizing the inputs for all neurons in layer $k+1$, $\sigma'(\mathbf{s}^{k+1})$ is the vector of corresponding balance values, and $\odot$ denotes term-wise multiplication.

Therefore, we have:
\begin{align}
\Delta w^k_{ij} &= c^{k+1}_i \sigma'(s^{k+1}_i)x^k_j 
= \left( (\dfrac{\partial \mathbf{x}^{k+2}}{ \partial \mathbf{x}^{k+1}})^T \mathbf{c}^{k+2}\right)_i \sigma'(s^{k+1}_i)x^k_j \\
&= \left( (\dfrac{\partial \mathbf{x}^{k+2}}{ \partial \mathbf{x}^{k+1}})^T (\dfrac{\partial \mathbf{x}^{k+3}}{ \partial \mathbf{x}^{k+2}})^T\mathbf{c}^{k+3}\right)_i \sigma'(s^{k+1}_i)x^k_j \\
&= \left( (\dfrac{\partial \mathbf{x}^{k+2}}{ \partial \mathbf{x}^{k+1}})^T (\dfrac{\partial \mathbf{x}^{k+3}}{ \partial \mathbf{x}^{k+2}})^T ...\; (\dfrac{\partial \mathbf{x}^m}{ \partial \mathbf{x}^{m-1}})^T \mathbf{c}^m\right)_i \sigma'(s^{k+1}_i)x^k_j \\
&= \left( (\mathbf{c}^m)^T \dfrac{\partial \mathbf{x}^m}{ \partial \mathbf{x}^{m-1}} \;...\; \frac{\partial \mathbf{x}^{k+3}}{ \partial \mathbf{x}^{k+2}} \frac{\partial \mathbf{x}^{k+2}}{ \partial \mathbf{x}^{k+1}} \right)_i \sigma'(s^{k+1}_i)x^k_j\\
&= \left( (\mathbf{c}^m)^T \dfrac{\partial \mathbf{x}^m}{ \partial \mathbf{x}^{m-1}} \;...\; \frac{\partial \mathbf{x}^{k+3}}{ \partial \mathbf{x}^{k+2}} \frac{\partial \mathbf{x}^{k+2}}{ \partial \mathbf{x}^{k+1}} \right)_i \frac{\partial x^{k+1}_i}{\partial w^k_{ij}}
\end{align}
If the activity of neurons in the output layer, driven by some external signal, is equal to the derivative of the loss $L$ with respect to the neural firing rate, i.e., $\mathbf{c}^m = -\frac{\partial L}{\partial \mathbf{x}^m}$, then we have:
\begin{align}
\Delta w^k_{ij} 
&= \left( -(\frac{\partial L}{\partial \mathbf{x}^m})^T \dfrac{\partial \mathbf{x}^m}{ \partial \mathbf{x}^{m-1}} \;...\; \frac{\partial \mathbf{x}^{k+3}}{ \partial \mathbf{x}^{k+2}} \frac{\partial \mathbf{x}^{k+2}}{ \partial \mathbf{x}^{k+1}} \right)_i \frac{\partial x^{k+1}_i}{\partial w^k_{ij}} \\
&=  -\frac{\partial L}{\partial x^{k+1}_i}\frac{\partial x^{k+1}_i}{\partial w^k_{ij}} = -\frac{\partial L}{\partial w^{k}_{ij}}
\end{align}
which is the defining weight update rule of backpropagation. 
\end{document}